\documentstyle[12pt]{article}
\newcommand{\BEQ}{\begin{equation}}  \newcommand{\EEQ}{\end{equation}}
\newcommand{\BAR}{\begin{array}}     \newcommand{\EAR}{\end{array}}
\newcommand{\BEA}{\begin{eqnarray}}  \newcommand{\EEA}{\end{eqnarray}}
\newcommand{\kv}{\frac{1}{4}}        \newcommand{\kh}{\frac{1}{2}}
\newcommand{\pv}{\frac{\pi}{4}}      \newcommand{\ph}{\frac{\pi}{2}}
 \newcommand{\cg}{C_\Ga}
 \newcommand{\al}{\alpha}
\newcommand{\kd}{{\textstyle \frac{3}{4}}}
\newcommand{\kpv}{{\textstyle \frac{\pi}{4}}}
\newcommand{\kpd}{{\textstyle \frac{\pi}{3}}}
\newcommand{\kph}{{\textstyle \frac{\pi}{2}}}
\newcommand{\kda}{{\textstyle \frac{3\pi}{8}}}
\newcommand{\kdv}{{\textstyle \frac{3\pi}{4}}}
\newcommand{\NNZ}{N=6,\;9,\;12}
\newcommand{\Es}{\mbox{\sf E}}       \newcommand{\hh}{\hspace*{2mm}}
\newcommand{\Ee}{\mbox{\sf E}_1}     \newcommand{\Ez}{\mbox{\sf E}_2}
\newcommand{\RR}{\mbox{{\sf R}\hspace*{-0.7em}{\sf I}\hspace*{3mm}}}
\newcommand{\CC}{\mbox{{\sf C}\hspace*{-0.40em}{\sf I}\hspace*{1mm}}}
\newcommand{\ZZ}{\mbox{{\sf Z}\hspace*{-0.8em}{\sf Z}}\hspace*{1mm}}
\newcommand{\Zd}{\;\mbox{{\ZZ}}_3}   \newcommand{\de}{\delta}
\newcommand{\Zn}{\;\mbox{{\ZZ}}_n}   
\newcommand{\ra}{\rightarrow}        \newcommand{\De}{\Delta}
        \newcommand{\vphi}{\varphi}
\newcommand{\fd}{\frac{\phi}{3}}     \newcommand{\pdr}{\frac{\varphi}{3}}
\newcommand{\la}{\lambda}            \newcommand{\si}{\sigma}
\newcommand{\om}{\omega}             \newcommand{\ome}{e^{2\pi i/3}}
           \newcommand{\Ga}{\Gamma}
\newcommand{\LV}{\langle v\mid}      \newcommand{\RV}{\mid v\rangle}
  \newcommand{\PX}{P-\frac{\phi}{3}}
 \newcommand{\PY}{P-\frac{2\phi}{3}}
\newcommand{\pp}{\frac{\pi-\varphi}{3}}   \newcommand{\Pmi}{P_{\min}}
             \newcommand{\NIF}{N\rightarrow\infty}
\newcommand{\hs}{\hspace*{1cm}}      
\newcommand{\hj}{\hspace*{3mm}}      \newcommand{\hi}{\hspace*{-3mm}}
\newcommand{\sui}{\phi=\vphi=\kph}   \newcommand{\hp}{\hspace*{6mm}}
\newcommand{\ny}{\nonumber}          \newcommand{\zpz}{2i\phi/3}
\newcommand{\zeile}[1]{\vskip #1 \baselineskip}
\newcommand{\RE}{\mbox{Re}}
\newcommand{\IM}{\mbox{Im}}

\begin{document}
\begin{titlepage}
\null    \begin{center}   \vskip 2cm
{\LARGE {\bf Excitation Spectrum and Correlation Functions of the\,\,
\mbox{${\ZZ\hspace{1mm}}_3$-Chiral} Potts Quantum Spin Chain}}
\vskip 1.5cm
{\large {\bf  A.\ Honecker$^*$ and G.\ von Gehlen$^+$}}  \zeile{1}
{\it Physikalisches Institut der Universit\"{a}t Bonn \\
Nussallee 12, 53115 Bonn, Germany} \\
\zeile{2}     \end{center} {\large {\bf Abstract :} }
We study the excitation spectrum and the correlation functions of the
$\;\ZZ_3$-chiral Potts model in the {\it massive high-temperature phase}
using perturbation expansions and numerical diagonalization. We are
mainly interested in results for general chiral angles
but we consider also the superintegrable case. For the parameter values
considered, we find that the band structure of the low-lying part of the
excitation spectrum has the form expected from a quasiparticle picture with
{\em two} fundamental particles.
\par Studying the $N$-dependence of the spectrum, we confirm the
stability of the second fundamental particle in a limited range of the
momentum, even when its energy becomes so high that it
lies very high up among the multiparticle scattering states. This is not
a phenomenon restricted to the superintegrable line.
\par  Calculating a non-translationally invariant correlation
function, we give evidence that it is oscillating. Within our numerical
accuracy we find a relation between the oscillation length and
the dip position of the momentum dispersion of the lightest particle which
seems to be quite independent of the chiral angles.
\vfill
\noindent   BONN-TH-94-20 \hspace{5mm} hep-th/9409170\hfill September 1994\\
$^*$ unp06b@ibm.rhrz.uni-bonn.de   \\
$^+$ unp02f@ibm.rhrz.uni-bonn.de        \end{titlepage}

\section{The $\;\;\ZZ_3-$chiral Potts model}
Recently, the $\hh\ZZ_n$-symmetrical chiral Potts model \cite{gri,auy} has
attracted particular interest \cite{acp}$-$\cite{hah}
\footnote{For a more complete list of references see e.g.\ \cite{kcoy}}
because of its special integrability properties and the rich structure
of its phase diagram. In this paper we consider the $\;\ZZ_3$-quantum chain
version of the model \cite{hkn}. The Hamiltonian of the model
contains complex coefficients and this gives rise to a large number of
ground state- (and also higher level-) crossings, and to the appearance
of both massive and massless oscillating phases.
We present evidence that the spectrum has a quasiparticle structure
all over the massive high-temperature phase. We also study the peculiar
momentum dependent effects which appear due to the parity-noninvariance of the
model: A quasiparticle may overlap with a two-particle spectrum for a
limited range of its momenta. We investigate whether by immersing
into the continuum the particle remains stable or ceases to exist as a
quasiparticle. Such an effect has been discovered recently in \cite{dkc}
for the superintegrable case of the chiral Potts chain. Finally, we present
finite-size calculations which establish directly an oscillatory behaviour
of certain correlation functions in the massive high-temperature phase.

\par The $\;\ZZ_3$-chiral Potts quantum chain with $N$ sites is defined by the
Hamiltonian:
\BEQ       H_N^{(3)} = -\frac{2}{\sqrt{3}}\sum_{j=1}^N
              \left\{ e^{-i\vphi/3}\si_j + e^{i\vphi/3}\si_j^+
  +\la \left(e^{-i\phi/3}\Ga_j\Ga_{j+1}^+ +e^{i\phi/3}\Ga_j^+\Ga_{j+1}
  \right)\right\}.                                       \label{H3}    \EEQ
The operators $\si_j$ and $\Ga_j$ act in a vector space $\CC^3$ located
at site $j$. They satisfy the relations     \BEQ
\si_i\Ga_j=\Ga_j\si_i\;\om^{\delta_{j,j'}};\hs\si_j^n=\Ga_j^n=1;\hp
\mbox{where}\hp \om=\ome. \label{dfs} \EEQ  We shall use the representation
\BEQ   \si_j = \left( \BAR{ccc} 1&0&0\\0&\om&0\\0&0&\om^2
       \EAR \right)_{\!j},  \hspace{1cm} \Ga_j =
 \left( \BAR{ccc} 0&0&1\\1&0&0\\0&1&0 \EAR \right)_{\!j}. \label{sG}    \EEQ
In this paper we will consider the ferromagnetic case $0 \leq \la < \infty$.
Using periodic boundary conditions
$\Ga_{N+1}=\Ga_1$, the Hamiltonian (\ref{H3}) commutes with the translation
operator $T$ and with the $\;\ZZ_3$-charge operator
$\hat{Q}=\prod_{j=1}^N\si_j$. Their eigenvalues have the form $e^{i P}$
and $\om^Q$, respectively, leading to conservation of the momentum
$P$ (which for finite chains takes the $N$ discrete values
$0\le 2\pi k/N< 2\pi$ with \mbox{$k=0,1,\ldots,N-1$}),
and to the conservation $mod\;\;3$ of the charge $Q=0,1,2$.
See e.g.\ \cite{gri,mcy} for the definition of the chiral Potts quantum chain
for general $\;\ZZ_n$ which for $n=2$ reduces to the standard Ising
quantum chain.                                         \par
The $\;\ZZ_3$-model contains three parameters: $\la$ which plays the r\^ole
of the inverse temperature, and $\phi,\; \vphi$ which are two chiral angles.
For
$\phi=\vphi=\ph$, the Hamiltonian (\ref{H3}) is `superintegrable', i.e.\
it fulfills the Onsager algebra \cite{gri,dav}. For
 \BEQ \cos{\vphi}=\la\cos{\phi}  \label{in} \EEQ
(\ref{H3}) has been shown to be integrable \cite{acp} due to a new
type of Yang-Baxter equation which contains two rapidities {\it not}
enjoying the usual difference property. For generic $\phi=\vphi$
the Hamiltonian (\ref{H3}) is self-dual but probably not integrable.
The self-dual model is parity-invariant only for $\phi=\vphi=0$. In this case
(\ref{H3}) is just
the standard ${\cal WA}_2$-symmetrical $\;\ZZ_3$-Potts quantum chain.
The phase diagram of (\ref{H3}) contains four distinct phases, two of
which are massless incommensurate phases. For recent reviews summarizing
the known properties of the model, see \cite{mcy,geh}.
\par  In this paper we shall study properties of the {\it massive
high-temperature phase} of the model which occurs for small $\la$:
$0<\la<\la_c(\phi,\vphi)$. The boundary of this phase, $\la_c(\phi,\vphi)$,
is known exactly only for a few special values of $\phi,\;\vphi$:
For $\phi=\vphi$, it is known that $\la_c(0,0) = 1$ (standard Potts
case), $\la_c(\ph,\ph)=0.901292\ldots$ (superintegrable case) \cite{acp}
and for $\phi\ra \pi\,,\; \vphi\ra \pi$ we have $\la_c\ra 0$ \cite{hkn}.
\par In Sec.~2 we start explaining our tools for analyzing the spectrum
of (\ref{H3}), summarizing known formulae and presenting simple
new perturbation formulae. In Sec.~3 we show that the finite-size data
follow the band structure expected for a quasiparticle spectrum. In
Sec.~4 we study the immersion of the $Q=2$-quasiparticle into the
many-particle continuum for a limited range of $P$.
In Sec.~5 we study a non-translationally invariant correlation function
and establish its oscillatory nature which previously had been inferred
only from wave-vector scaling arguments. A simple relation between the dip
position of the $Q=1$-particle dispersion relation and the oscillation
length is found to be approximately valid over a large parameter
range. Finally, Sec.~6 presents our Conclusions.

\section{Methods of calculation}
The complete analytic calculation of the spectrum of (\ref{H3}) in the
superintegrable case $\phi=\vphi=\ph$ by the Stony-Brook group (\cite{dkc}
and references therein) has established that for these parameter values the
spectrum has a quasiparticle structure, i.e.\ in \cite{dkc} it has been shown
that in the large chain limit {\em all} excitations
$\De E_{Q,r}(P,\la)$ above the ground state are composed of two
fundamental excitations $\Ee(P)$ and $\Ez(P)$
according to the following rules:                         \BEQ
\Delta E_{Q,r}(P,\la)=\sum_{k=1}^{m_r}{\sf E}_{Q^{(k)}}(P^{(k)})\, ,
  \hj  P = \sum_{k=1}^{m_r} P^{(k)} \hbox{ mod } 2 \pi \, , \hj
  Q = \sum_{k=1}^{m_r} Q^{(k)} \hbox{ mod } 3 .  \label{QP} \EEQ
Additionally, the fundamental quasiparticles satisfy a Pauli principle, i.e.\
\BEQ Q^{(i)}=Q^{(j)}\hp\mbox{implies}\hp P^{(i)}\ne P^{(j)}.\label{Pau}\EEQ
\par  For more general $\phi,\; \vphi$ and
general $\;\ZZ_n$ numerical results were presented in \cite{gho} which
suggested that such a quasiparticle structure (with $n-1$ quasiparticles
in the $\ZZ_n$-case) is present all over the massive high-temperature phase.
Here we want to give further evidence for
such a structure for the $\ZZ_3$ model (\ref{H3}) for quite general $\phi$,
$\vphi$. For this we compare multi-particle levels to the basic
single-particle levels and check whether the former can be composed out of
of the latter ones. Single-particle levels are expected to show exponential
convergence in $N$ to the thermodynamic limit, while composite states
should show power convergence with the power indicating the number of
constituents \cite{geh,gho}.
\par In order to analyze the spectrum, three different approaches will
be used: \par 1) In each charge-momentum sector we calculate numerically
the lowest $\geq 10$ energy levels of (\ref{H3}). For chains of length
$N\le 8$ complete numerical diagonalization is easily possible, while
for up to $N\le 12$ sites we use Lanczos techniques. This way, both
the basic quasiparticle dispersion relations $\Ee(P)$ and $\Ez(P)$ and
many continuum levels can be obtained. The limit $\NIF$ may be calculated
using standard finite-size extrapolation techniques.
In general, the convergence in $N$ depends on the mass scale involved.
For small $\la$ the mass scale is large but also for $\la$ close to $\la=1$,
where the mass scale is small, good information can be obtained
as will be shown below.
\par 2) Compact analytic, though approximate formulae for each {\em lowest}
level in the $Q\neq 0$ sectors can be calculated for general $\phi$
and $-\frac{\pi}{2} <\vphi < \frac{\pi}{2}$ using perturbation theory
in $\la$. We find up to order $\la^3$:     \BEA
\Ee(P)&:=&\De E_{1,0}(P,\phi,\vphi)\ny\\&=&4\sin{(\frac{\pi-\vphi}{3})}
    -\frac{4\la}{\sqrt{3}}\left\{{\cal F}(P)
    +\al-2\beta-(\al^2+2\beta^2)\cos{\phi}\right\}
    +{\cal O}(\la^4),   \label{ee}    \EEA
with \BEQ {\cal F}(P)=(1-A)\cos{p}+\al\cos{(p+\phi)}+\beta\cos{(2p)}
 + B\cos{(2p+\phi)}+2\beta^2\cos{(3p)},  \EEQ
where \BEQ  p=\PX, \hs \al= \frac{\la}{6\cos{(\pp)}},\hs
           \beta=\frac{\la}{6\cos{(\pdr)}},   \EEQ  and
$A=3\beta^2-2\al\beta+2\al^2;\;\; B=2\beta^2+2\al\beta-\al^2$. In the
 superintegrable case we have $\al=\beta=\la/(3\sqrt{3})$ and $A=B=\la^2/9$.
\par
For $0 \le \vphi< \ph$ the lowest $Q=2$ excitation is given by:
\BEQ  \Ez(P) :=\De E_{2,0}(P,\phi,\vphi) = \De E_{1,0}(P,-\phi,-\vphi).
                                           \label{ez}    \EEQ
These formulae are expected to be valid for $\la$ not too large, i.e.\ if the
distance of the level from other multiparticle levels is not too small.
Of course, this nondegenerate perturbation calculation is useless once
a level penetrates into the continuum of multiparticle states. We shall
see that there are parameter values for which
 this happens in the $Q=2$-sector.
\par The $P$-dependence of $\Ee$ is contained in ${\cal F}(P)$ only. So,
e.g.\ to find the value of the momentum $P=\Pmi$, for which the lowest
energy is reached (because of parity violation this is not at $P=0$),
we have to look for the solution of $\partial {\cal F}/\partial p=0$.
For non-zero $\phi,\;\cos{\pp}$ and $\cos{\pdr}$, we see that
 $p_{\min}\equiv \Pmi-\fd$ is of order $\la$. So, expanding we get
\BEQ \Pmi=\fd-\al\sin{\phi}+{\cal O}(\la^2). \label{PMM}\EEQ Observe that,
in order to obtain $\Pmi$ to order $\la$, we have to use
the second order expansion of $\Ee(P)$. Starting for $\la\ll 1$
at $\Pmi=\fd$, the position of the minimum shifts to smaller values of $P$
for increasing $\la$. Using (\ref{ez}), one
obtains immediately the minimum of the curve $\Ez(P)$.
\par 3) For the superintegrable case $\sui$ analytic formulae are
available from \cite{dkc} for the whole spectrum. We shall use here only
the formulae for the energies of the two fundamental excitations
which are given by
\BEA \hspace*{-9mm}\Ee(P_r)=e_r\hi&=&\hi 2\mid 1-\la \mid + \frac{3}{\pi}
\int_1^{\mid\frac{1+\la}{1-\la}\mid^{2/3}}\!\!\! {\rm d}t\;\;
\frac{v_r (2 v_r t + 1)}{v_r^2 t^2 + v_r t + 1}
\sqrt{ \frac{4\la}{t^3-1} -(1-\la)^2}      \ny  \\      \hspace*{-6mm}
       \Ez(P_{2s}) = e_{2s} \hi&=&\hi 4\mid 1-\la \mid + \frac{3}{\pi}
\int_1^{\mid\frac{1+\la}{1-\la}\mid^{2/3}}\!\!\!{\rm d}t\;\;
\frac{v_{2s} (4 v_{2s}^2 t^2-v_{2s}t + 1)}{v_{2s}^3 t^3 + 1}
\sqrt{ \frac{4\la}{t^3-1} -(1-\la)^2},      \label{Sz}     \EEA
where the auxiliary variables $v_r$ and $v_{2s}$ are related to the momenta
by:
\BEQ v(P)=+\sqrt{ \frac{1-\cos{P}}{1-\cos{(P+\frac{2\pi}{3})}}} \, ,
  \label{vv} \EEQ
\BEQ v_{2s}(P) = -v(2\pi-P) \, , \qquad
     v_r(P)= \left\{\BAR{cc}
                      -v(P) & \mbox{ for } 0\le P\le 4\pi/3; \\
                      +v(P) & \mbox{ for } 4\pi/3\le P\le 2\pi.
             \EAR\right.
 \label{vrs}  \EEQ
The integral appearing in $\Ez(P_{2s})$ has a singularity at $P=2\pi/3$, so
that the $Q=2$-quasiparticle level is defined only for $2\pi/3 \le P\le 2\pi$.
Except for this singularity, there is no problem to evaluate the
integrals in (\ref{Sz}) numerically: using $\xi=\sqrt{t-1}$ instead of $t$
 as our integration variable, the integrands are smooth at $\xi=0$ or $t=1$.
\par Comparing these different methods, we can keep control about the
 precision of the values to be used for $\Ee(P)$ and $\Ez(P)$ for all
 $\phi, \vphi$ and $\la<1$.
We shall comment on such comparisons as we go on presenting our results.
Methods 1) and 2) can be applied equally well also in the non-integrable
and in the non-hermitian situation (the Lanczos method can also be adapted
to the case of non-hermitian matrices \cite{yil}) which arises for
 complex $\phi$ or $\vphi$ e.g.\ if (\ref{in}) is imposed.
 This way we can cover a large range of $\phi$ and $\vphi$.
\par As a first example, from which one can judge the usefulness of the
different methods, in Fig.~1 we show the lowest part of the $Q=1$-spectrum
for the superintegrable case $\sui$ at the large value of $\la=0.90$ which
is close to the phase transition point.
We see that the finite-size data for $N=12$ lie still within $<6\%$ on
the exact curve (\ref{Sz}). For small $P$, even the points for small $N$ move
nicely along the exact curve. Of course, with only $N\le 12$ we
cannot get directly points in the interval $0<P<\pi/6$. Considering the
large value of $\la$ and the vicinity of the phase transition, it is
remarkable that the $3rd$-order perturbation curve remains
close to the exact curve, especially for $P>1.2\pi$.
It mainly fails to describe details of the dip at small $P$, and so is
not useful for the determination of the
transition to the incommensurate (IC)-phase (where the ground state looses
translational invariance).

\section{Quasiparticle- and band-structure of the spectrum}
In order to investigate the band structure expected from the composition
of the two quasiparticles, in Fig.~2 we start considering the simple
situation which occurs for a small value of $\la$, here in particular
$\la=0.1$, and a self-dual choice of the chiral
angles $\phi=\vphi=\kpv$ well below the superintegrable line.
The single-particle dispersion curves can be determined with high precision:
the $N=12$ finite-size value for $\Ee(P)$ agrees to better than
$10^{-4}$ with the value given by formula (\ref{ee}). Also the agreement in
$\Ez(P)$ between $N=12$ and perturbation theory is better than
$10^{-3}$ over the whole range of $P$. Both curves are rather flat,
corresponding to a large mass in a Klein-Gordon-type approximation to the
dispersion curve. Of course, at $\la=0$ the $P$-dependence is completely
absent, as is immediately seen from the perturbation expansion.
\par From these single-particle dispersion curves we can calculate the shape
of the two-particle bands in all three charge $Q$-sectors.
These bands are bounded by $\Es_Q^\pm(P),$ given by
\BEQ     \Es_Q^+(P)= \max_{-\pi<q\le\pi}\;\left\{\Es_{Q_1}(P/2+q)+
\Es_{Q_2}(P/2-q)\right\}, \hs  Q=Q_1 + Q_2 \mbox{ mod 3} \label{ban} \EEQ
(and similarly $\Es_Q^-(P)$ with $\max_q$ replaced by $\min_q$).
Now, as long as the $P$-dependence of the quasiparticle dispersion curves has
the form \BEQ \Es_{Q_i}(P)=A_i+\sum_j B_{ij}\cos{(P-\al_{ij})},\label{Dk}\EEQ
the boundaries $\Es_Q^\pm(P)$ for two {\em equal} particles are obtained for
$q=0,\pi$, respectively, and we have simply
\BEQ \Es_Q^\pm(P)=\Es_{Q_1}(P/2)+\Es_{Q_2}(P/2). \label{Ej}\EEQ
Of course, we have to use the doubled Brillouin zone for $P$.
In the perturbation formula (\ref{ee}) the first term to spoil the form
(\ref{Dk}) is the term $\sim \la^2\cos{(2\PY)}$. For the parameters chosen
in Fig.~2 this is quite small. For two charge conjugate particles
(in our $\;\ZZ_3$-case this occurs for $Q=0$), with
$\alpha_{1,j}=-\al_{2j}$ and $A_1=A_2$, $B_{1j}=B_{2j}$ (which is true
in order $\la$ of (\ref{ee}), (\ref{ez})), we get
\BEQ    \Es_1(P/2+q)+\Es_2(P/2-q)=2\left\{A_1+
   \sum_j B_{1,j}\cos{(P/2)}\cos{(q-\al_{ij})}\right\}. \label{Ek}  \EEQ
so that the band width becomes zero at $P=\pi$.
The $\la^2$-contributions no longer obey fully (\ref{Dk}) with the effect
that the lower corner of the zero-width point is rounded off.

\par In Fig.~2 we have plotted the nine lowest $N=12$-sites levels
of each charge and each momentum state. As expected in the quasiparticle
picture, we see that at all available values of $P\neq\pi$ there are three
bunches of levels with the correct charge filling up and following the
respective two-particles band ranges. The $P$-values, where the respective
bands have zero width is nicely seen to be at $P=\pi\pm 2\Pmi\approx \pi\pm
\pi/6$ for the $Q=2$- and $Q=1$-bands, respectively, and at $P=\pi$ for
the $Q=0$-band. For $Q=1$ and $Q=2$, according to whether $k$ is odd or even,
the uppermost three, respective two highest levels start forming the
next bands which at these energies are expected from three-particles states.
\par Only for $P=\pi$ four $Q=0-$points appear outside of the bands. The
two points slightly above $E=8$ are within the expected three-particle band
but the two points above and below $E=6.5$ must be two-particle states
which, because of power convergence at $N=12$, are still out of the here
very narrow band. Such finite-size effects have been discussed in detail for
$P=0$ in \cite{gho}. There we found that, taking finite-size effects
into account, for $P=0$ the quasiparticle picture explains the data very well.
\par In Fig.~3 we show the analogous picture for a somewhat larger value of
the inverse temperature, $\la=0.25$, and a (again self-dual) choice of
the chiral angles $\phi=\vphi=67.5^\circ$, closer to the superintegrable
value $\phi=\vphi=90^\circ$.
For the $Q=1$-ground state the third order perturbation formula
still agrees within $\approx 10^{-3}$ over the whole range of the momentum
with the $N=12$ sites data, however, there are somewhat larger
($\approx 5 \cdot 10^{-2}$) discrepancies for the lowest $Q=2$-level.
 \par The band structure in all three charge sectors
is of the same shape as before and 
all higher
levels lie within the bands (this time we show {\em all} higher levels
for $N=4,\ldots 12$ sites but draw only the boundary curves for the $Q=2$-
band). The approximation (\ref{Ej}) for the {\em lower} bound of the
$Q=2$-two particle band is no longer perfect, it is interesting
to see that the finite-size data at the most narrow point $P\approx 1.22\pi$
do follow the curvature expected. With increasing $N$ the bands are filled
densely from the {\em interior} which is consistent with but of course
no proof of the presence of the Pauli principle effect (\ref{Pau}).
\par More important is to observe in this Figure that, due to the parity
non-invariance, the maximum of the $\Ez$-curve is shifted to the left of
$P=\pi$, while the two-$Q\!=\!1$-particle band is shifted to the right of
$P=\pi$. This
causes the curve for $\Ez$ to penetrate into the two-particle band.
This may make the $Q=2$-quasiparticle unstable in a limited range of the
momentum around $P\approx 2\pi/3$. Indeed, the convergence with $N$ of the
lowest level in the $Q=2$-sector deteriorates considerably at these
momenta. \par In \cite{geh,gho} it has been pointed out that the
$Q=2$-particle enters the continuum composed out of two $Q=1$-particles
at $P=0$ on the superintegrable line . However, we see that
 for $P\approx 2\pi/3$ this
effect appears already for lower values of $\phi$. With increasing $\la$ this
"threshold" moves to smaller values of $\phi$. It is tempting to
suspect that the limitation of the range of the momentum for the
$Q=2$-quasiparticle discovered in \cite{dkc} for the superintegrable case,
see eq.~(\ref{Sz}),  might be due to this penetration effect. In the following
Section we try to shed some light on this problem by studying the
chain-size-dependence of the spectrum.
\par Not surprising in view of the penetration of the $Q=2$-quasiparticle
into the continuum, the perturbation formula for $\Ez(P)$ of
eqs.~(\ref{ee}), (\ref{ez}), which for $\la=0.25$ and
$\phi=\vphi=67.5^\circ$ was still good, becomes
quite useless for larger values of $\la$ and $\phi$: E.g.\ for $\la=0.5$
and $\phi=\vphi=60^\circ$ it is off by $20-25\%$ around $P=0$, while in the
backward direction it describes the finite-size data still within $<3\%$.
Going at $\la=0.5$ to $\phi=\vphi=75^\circ$, in the forward direction the
$\la^3$-perturbation formula value is already too large by about a
factor of three (only around $1.2\pi\le P\le 1.5\pi$  the curve comes back
to the finite-size data within a few percent).
\par Looking again into Fig.~1 we see that there are ten points which appear
well below the curve $\Ee(P)$. This happens because close to $\la=1$ also the
$Q=1-$quasiparticle enters a multiparticle band. This time the band is
formed by four of the same $Q=1$-particles as can be seen by estimating
its lower boundary from $\Ee(P-3P_{\min}) +3\Ee(P_{\min})$. So it is not
trivial that the perturbation formula for $\Ee(P)$ remains useful around
$\la\approx 0.9$ as seen in Fig.~1.

\section{Stability of the $Q=2$-quasiparticle in the two-particle continuum}
\subsection{The superintegrable case}
As we have seen, the description of the $Q=2$-quasiparticle by perturbation
theory breaks down as the chiral angles approach the superintegrable value
$\sui$. This, of course, is related to the overlapping of the $Q=2$-particle
with the continuum. {\em At} $\sui$, where for $P=0$ the $Q=2$-particle sits
just at the threshold of the two $Q=1$-particle scattering states (sometimes
expressed in the form that for these chiral angles $m_2=2m_1$ \cite{geh}).
Since one can guess the
correct diagonal linear combination of states at this point for $0\le\la<1$,
degenerate perturbation theory is possible here \cite{per}. However, it
is not known how to generalize this at $\sui$ to $P\neq 0$, because in
the standard basis the diagonal combination of states probably is highly
complicated.
\par Finite-size methods are not affected by these problems and
still give good information (these will be useful even for
$\vphi>\ph,\;\phi>\ph$) as we shall see by comparison to the exact $\NIF$
formula (\ref{Sz}). In particular, we are going to check now
whether we can confirm the details of the peculiar energy-momentum rule of
the $Q=2$-quasiparticle (\ref{Sz}) by finite-size calculations. After showing
that this is possible, we go ahead and look whether a similar unusual
energy-momentum rule is also found for non-superintegrable chiral angles.
The question whether integrability plays a special r\^ole here will be studied
choosing chiral angles for which the system probably is not integrable.
\par Fig.~4 shows the curves calculated from (\ref{Sz}) together with
the 9 lowest $N=12$ finite-size eigenvalues for each $Q,\;P$ for $\sui$
and $\la=0.50$ which is right in the center of the high-temperature phase.
\par In the lower part of Fig.~4 we see the curve $e_r(P)$ drawn according
to the analytic formula (\ref{Sz}). The $Q=1,\;N=12$-data (crosses) lie
within $<2 \cdot 10^{-5}$ on this curve \footnote{Also the $\Ee(P)$-third
order perturbation curve (\ref{ee}) (not shown in the Figure) agrees still
within $<2\%$ with the exact expression. Observe that the $\la$ here is much
smaller than the $\la=0.9$ chosen in Fig.~1. Compare also the fourth columns
in Tables~1 and 2 below.}.
\par  The small dashed
lines mark the two-particle band expected for two $Q=1$-quasi\-par\-ticles
according to both (\ref{ban}) (rounded lower curve at $P\approx 1.3\pi$) and
the approximation (\ref{Ej}) (crossing curves). For $P< 3\pi/2$ the
$Q=2$-particle dips into the two-particle band, which for finite-$N$
(consider only the $Q=2$-squares in the Figure) contains many states.
Studying their $N$-dependence one finds that in general these points move
considerably when $N$ is varied, indicating an only power-convergence to
the thermodynamic limit. However, one notices that $e_{2s}$ neatly {\em
passes through} several $N=12$-finite-size eigenvalues which for
$P\le 4\pi/3$ are some of the higher lying levels. We are now going to argue
that it is no random coincidence that the $e_{2s}-$curve passes through the
$N=12$-points but this comes because these states belong to an
exponentially converging sequence, characteristic of a fundamental
quasiparticle (For the concept of a quasiparticle see the beginning of Sec.~2).
\par Exponential convergence for states of fixed momentum $P$ can be inferred
from finite-size data directly only if a sequence of different $N$-values
at the same $P$ is accessible. Since $P=2\pi k/N$, with $k$ integer, this is
possible e.g.\ for $P=\pi$, where we can use data for all even $N$.

\begin{table} \vskip 5mm  \renewcommand{\arraystretch}{1.1} \noindent
\begin{tabular}{|c||c|cl|c||c|cl|} \hline \multicolumn{1}{|c||}{ }&
\multicolumn{4}{c||}{$Q=1$}&\multicolumn{3}{c|}{$Q=2$} \\  \hline
$P$ &$e_r$ & \multicolumn{2}{c|}{Finite chain} & Perturb.&
 $e_{2s}$&\multicolumn{2}{|c|}{Finite chain
$\;$/$\;$level \#} \\ \hline
$\frac{\pi}{6} $ &0.8659401&0.8659349&$N=12$&0.855991&      &      & \\
$\frac{\pi}{3} $ &1.1875666&1.1875601&$N=12$&1.157407&      &      & \\
$\frac{\pi}{2} $ &1.7400548&1.7400475&$N=12$&1.741481&      &      & \\
$\frac{2\pi}{3}$ &2.3110674&2.3110599&$N=12$&2.333333&6.0000000&5.9999978&
$N=12/8th$\\  &  &2.3108898&$N=9$&  &  &5.9998044 & $N=9/5th$\\
              &  &2.3099686&$N=6$&  &  &5.9963767 & $N=6/4th$\\
$\frac{5\pi}{6}$
&2.7776630&2.7776545&$N=12$& 2.784303&5.7768864&5.7725860&$N=12/7th$\\
$\pi$ &3.0691561&3.0691471&$N=12$&3.064815&5.4748476&5.4754941 &$N=12/7th$\\
           &  &3.0690841&$N=10$&  &  &5.4727480 & $N=10/5th$ \\
           &  &3.0687830&$N=8$&   &  &5.4797166 & $N=8/5th$  \\
$\frac{7\pi}{6}$
&3.1472946&3.1472849&$N=12$&3.144009&5.0141793&5.0139793&$N=12/6th$\\
$\frac{4\pi}{3}$
&3.0000000&2.9999893&$N=12$&2.990741&4.3971362&4.3972296&$N=12/5th$\\
           &  &2.9998304&$N=9$&  &  &4.3960055 & $N=9/4th$ \\
           &  &3.0002000&$N=6$&  &  &4.3984911 & $N=6/3rd$ \\
$\frac{3\pi}{2}$
&2.6408227&2.6408101&$N=12$&2.628890&3.6717090&3.6216780&$N=12/1st$\\
       &  &2.6405539&$N=8$&  &  &3.6716026 & $N=8/2nd$ \\
$\frac{5\pi}{3}$
&2.1133290&2.1133098&$N=12$&2.111111&2.9355387&2.9353258&$N=12/1st$\\
       &  &2.1176771&$N=6$&   &  &2.9386269 & $N=6/1st$ \\
$\frac{11\pi}{6}$
&1.5070578&1.5070792&$N=12$& 1.511994&2.3525974&2.3572572&$N=12/1st$\\
$2\pi$ &1.0000000&0.9999978&$N=12$&1.009260&2.0000000&1.9999893&$N=12/1st$\\
         &   &0.9999851&$N=11$&  &  &1.9999701&$N=11/1st$\\
         &   &0.9999405&$N=10$&  &  &1.9999247&$N=10/1st$\\
\hline  \end{tabular}  \vskip 5mm  \par \noindent Table 1:
Comparison of the exact formulae (\ref{Sz}) with finite chain size data for
$N=12$ and less sites and with the $3rd$-order perturbation formula for
$\la=0.50$ and $\sui$ (superintegrable case). In the last column we indicate
how high up in its $(Q=2, P)$-sector the quoted energy level appears. For
$Q=1$ the finite-$N$-values are always the $Q=1$-ground state ($1st$) levels.
\vskip 5mm \end{table}
However, being limited to $N\le 12$,
there are only a few other values of $P$ where more than one $N$ can be used,
e.g.\ $P=2\pi/3$ which is reached for $N=12,\;9,\;6,\;3$
with $k=4,\;3,\;2,\;1$. Table~1 gives such data for the case $\la=0.5$ of
of Fig.~4. We compare the convergence in $N$ of the high-lying $Q=2$-levels
which are very close to the curve $e_{2s}$ with the lowest $Q=1$-data. About
the $Q=1$-levels we are on quite safe grounds to assume that these approximate
the well-isolated $Q=1$-quasiparticle exponentially fast for $\NIF$. Table~1
shows clearly that the convergence in $N$ to the high-lying $Q=2$-levels is
practically as good as that of the $Q=1$-levels and consequently also
exponential. This way we can trace the
$Q=2$-quasiparticle, if it is present, over its whole range of the momentum
by our finite-size methods. Although at other values of
$P$ (e.g.\ $P=7\pi/6$) we have only one "measurement" for $N=12$, we conclude
from the closeness of the $N=12$-value to the exact result that also here
we probably have exponential convergence. In contrast, in the case of power
convergence, typically the $N=12$-values are off the $\NIF$-limit still by
several percent.

\begin{table}
\vskip 5mm    \noindent
\begin{tabular}{|c||c|cl|c||c|cl|} \hline \multicolumn{1}{|c||}{ }&
\multicolumn{4}{c||}{$Q=1$}&\multicolumn{3}{c|}{$Q=2$} \\  \hline
$P$ &$e_r$ & \multicolumn{2}{c|}{Finite chain} & Perturb.&
 $e_{2s}$&\multicolumn{2}{c|}{Finite chain
$\;$/$\;$level \#} \\ \hline
$\frac{\pi}{6}$&0.224478&0.226286&$N=12$&$\hspace*{-2mm}-0.0161071$& & & \\
$\frac{\pi}{3} $ &1.064266&1.069278&$N=12$& 0.8840000&    &      & \\
$\frac{\pi}{2} $ &2.016074&2.022656&$N=12$&2.1927771&     &      & \\
$\frac{2\pi}{3}$ &2.882061&2.889695&$N=12$&3.0800000&7.600000&$7.5955496\!\!$
&$N=12/26th$\\ & &2.915791&$N=9$&  &  &7.595578& $N=9/11th$\\
                        &  &3.003151&$N=6$&  &  &7.608462& $N=6/5th$\\
$\frac{5\pi}{6}$
&3.551636&3.560115&$N=12$&3.5107612&7.251272&7.237896&$
  N=12/21th$\\
$\pi   $ &3.952458&3.961730&$N=12$&3.8360000&6.797425&6.7850878$\!\!$&
$N=12/17th$\\ & &3.977025&$N=10$& & &6.778859 & $N=10/10th$ \\
                         &  &4.010626&$N=8$&  &  &6.795053 & $N=8/7th$  \\
$\frac{7\pi}{6}$
&4.040930&4.051059&$N=12$&4.0161071&6.114959&6.100765&$N=12/14th$\\
$\frac{4\pi}{3}$
&3.800000&3.811210&$N=12$&3.8000000&5.190112&5.176061&$N=12/6th$\\
           &  &3.843411&$N=9$&  &  &5.173303 & $N=9/4th$ \\
           &  &3.950279&$N=6$&  &  &5.218181 & $N=6/3rd$ \\
$\frac{3\pi}{2}$ &3.238310&3.251180&$N=12$&3.2112229&4.056564&4.041591&
$N=12/3rd$\\  &  &3.310811&$N=8$&  &  &4.050302 & $N=8/3rd$ \\
$\frac{5\pi}{3}$
&2.389589&2.405991&$N=12$&2.3600000&2.777609&2.763267&$N=12/2nd$\\
              &  &2.592734&$N=6$&  &  &2.804678 & $N=6/1st$ \\
$\frac{11\pi}{6}$
&1.316212&1.352448&$N=12$&1.2452388&1.450685&1.438850&$N=12/1st$\\
$2\pi $ &0.200000&0.187664&$N=12$&0.2000000&0.400000&0.411210&$N=12/1st$\\
         &   &0.186542&$N=11$&  &  &0.418118&$N=11/1st$\\
         &   &0.186060&$N=10$&  &  &0.428335&$N=10/1st$\\
\hline  \end{tabular}  \vskip 5mm  \par \noindent            Table 2:
Comparison of the exact formulae (\ref{Sz}) with finite chain size data for
\mbox{$N\le 12$} sites and with the $3rd$-order perturbation formula for
$\la=0.90$ and $\sui$ (superintegrable case). In the last column we indicate
how high up in its $(Q=2, P)$-sector the particular quoted energy level
appears. In contrast to Table~1, here for $Q=1$
the levels approximating the curve $\Ee(P)$ are not always the
$Q=1$-ground state levels. Recall Fig.~1 which showed that the
$Q=1-$quasiparticle is entering the continuum for $2\pi/3\le P\le \pi$.
\vskip 5mm    \end{table}
\par We like to emphasize that the choice of levels, even if we get very high
up in the spectrum, in practically all cases is quite unambiguous: e.g.\ for
$\la=0.5$ and $P=\frac{2\pi}{3}$ the next neighbouring levels to those
quoted in Table~1 are: for $N=12$: $5.8192 / 6.0000 / 6.1965$, for $N=9$:
$5.6962 / 5.9998 / 6.2942$ and for $N=6$: $5.3865 / 5.9964 / 7.8306$.
For $\la=0.5$ and $P=\pi$ at $N=12$ we have the neighbours $5.2154 / 5.4755 /
5.9303$ which are also well separated. That we are not victims of
coincidences will be very clear from Fig.~5 below and also from Table~2
which gives the analogous results for $\la=0.90$. Here, since we are much
closer to the phase transition line, the level structure is much more
involved. Nevertheless, although the mass scale of the system is smaller
by a factor 5 as compared to $\la=0.5$ (as a relevant scale we may consider
$\Ee(P=0)=2(1-\la)$), and we have to go up to the $26th$ level at $P=2\pi/3$
and $N=12$, the convergence is still excellent.
\par Notice that the convergence is not monotonous in $N$ but with
oscillations superimposed, a phenomenon quite common in the chiral Potts
model \cite{geh}.             \par To show that
{\em below} $P=2\pi/3$ there is {\em no} exponentially converging
level is not so easy because we do not know precisely where to search.
We cannot search in the immediate neighbourhood below $P=2\pi/3$, because
for $N\le 13$ sites (with some effort, $N=13$-chains can be calculated but
are quite useless here since $13$ is a prime number), the next lowest point
where we can test for fast convergence is $P=\pi/2$. There we do not see any
convincingly close levels in the $N=12,\;k=3$ and $N=8,\;k=2$-spectra.
A better argument that $P=2\pi/3$ is the lower endpoint of the
$Q=2$-quasiparticle with respect to $P$ will be given below in connection
with Fig.~4, using the change of the spectrum at $P=2\pi/3$
with $\phi,\;\vphi$ when moving away from the superintegrable line.
\par Summarizing, we see that even from our numerical studies one
obtains strong evidence that one can interpret the spectrum of (\ref{H3})
on the superintegrable line in terms of two elementary quasiparticles
with energies $e_r(P)$ and $e_{2s}(P)$. The unusual features of the
$Q=2$ quasiparticle are clearly reflected in the finite-size data.
Decreasing $P$ from $2 \pi$ towards lower values,
the $Q=2$ excitation crosses the scattering threshold of two $Q=1$
excitations at some point but stays stable.

\subsection{The non-superintegrable case}
\par We expect that the limitation of the $Q=2$-quasiparticle to only
part of the range of the momentum should not only appear at the superintegrable
line $\sui$ but by continuity should be valid, maybe with changing range,
also for neighbouring values of $\phi$ and $\vphi$.
\par From our earlier calculations in \cite{gho} we get a hint about the
endpoint at $P=2\pi$: For $\phi=\vphi>\ph$ this endpoint should move to
$P<2\pi$ because we found that the $Q=2-$particle becomes unstable at
$P=0$ (or $P=2\pi$) (which was the only value for $P$ considered there).
\par In order to get more information, we now study the $Q=2$ spectra
at $P=2\pi/3$ for $\NNZ$ sites as functions of the chiral angle
$\phi$ at fixed $\la=0.50$. In order to vary only one parameter, we have
used both the self-dual choice $\phi=\vphi$ and the integrable choice
(\ref{in}) but we found that both behave very similarly. So here we
present only results for the self-dual case. For this choice (\ref{H3})
probably is not integrable, and the effects we are going to see should not
 be due to the integrability.
\par In Fig.~5 one clearly distinguishes three curves, one for each
$\NNZ$, which follow each other very closely for $\phi\ge 90^\circ$ and
separate in the region $\phi<90^\circ$. {\em At} $\phi=90^\circ$ this is
the $Q=2-$particle of (\ref{Sz}), and the three values for $\NNZ$ are those
of Table~1 ($5.9999978\;\;etc.$). Notice that indeed there are no other
levels close to 6.000 around $\phi=90^\circ$, and that nowhere else in the
picture are lines from $\NNZ$ following closely the same course. We see that
the $Q=2$-quasiparticle curve can be followed uniquely up to $\phi=125^\circ$,
and that the fast convergence from $N=6$ to $N=12$ is still there.
Around $\phi\approx 120^\circ$
the convergence is quite perfect. The energy values at $\phi=120^\circ$
being $5.57251;\;5.57211;\;5.57192$ for $\NNZ$, respectively.
\par For larger angles the $Q=2$-quasiparticle level gets so high up in
the spectrum that computational problems limit to follow it for $N=12$:
at $\phi=126^\circ$ it is already the $30th$ level in its sector. In
Fig.~5 we have plotted up to the $29th$ $N=12$-level in the $Q=2$,
$P=2\pi/3$-sector.
Around $\phi\approx 100^\circ$ the convergence is somewhat worse but still
quite good: $5.81200;\;5.79749;\;5.79150$ at $\phi=99^\circ$. We interpret
this as continuing presence and stability of the $Q=2-$quasiparticle for
$90^\circ\le\phi<125^\circ$ at $P=2\pi/3$.
\par However, for $\phi<90^\circ$ the curves for $\NNZ$ clearly separate.
We consider this to be an indication that here the $Q=2-$particle gets out
of its stability range. This is consistent with the analytic result that for
$\sui$, $P=2\pi/3$ is the lower momentum bound of the second quasiparticle.
\par On the scale on which Fig.~5 is drawn it is difficult to show that
apparent crossings of level-lines with the same $N,\; Q$ and $P$ are in fact
avoided crossings, since the levels come very close to each other.
E.g.\ consider the $N=9$-levels which around $\phi\approx 99.035^\circ$ seem
to cross at $E\approx 5.797$. At their closest encounter they keep separated
in energy by $\Delta E\approx 0.0007911$.
\par It remains to see what happens above the superintegrable line for
other momenta. E.g.\ for $P=\pi$ we find that the three $Q=2$-levels for
 $N=8,\;10,\;12$, which were very close to each other for $\sui$,
stay very close together if we move in $\phi$. Table~3 gives a few numbers
for the energy of the $Q=2$-particle obtained by following this triple.
At intermediate momentum values, where
we have only one $N\le 12$ available, we can make use of a property which
is clearly seen in Fig.~5 but which is qualitatively the same for other
$P$, too: the levels approximating the $Q=2$-quasiparticle have a slope
$\partial \Ez/\partial \phi$ which is very different from that of other
levels. Using this feature we can follow the quasiparticle line
to larger $\phi$, starting from the known value at $\sui$. Of course, when
using just this method we have no direct control over the size of errors.

\begin{table}
\vskip 5mm    \begin{tabular}{|c|cccccc|} \hline
$\phi = \vphi$ & $90^\circ$ & $96^\circ$ & $102^\circ$ & $108^\circ$
    & $114^\circ$ & $120^\circ$ \\ \hline
$P = 2 \pi/3$ & $6.0000(1)$ & $5.84(1)$   & $5.745(5)$  & $5.687(7)$   &
                                  $5.626(2)$   & $5.5718(3)$  \\
$P = \pi$     & $5.475(1)$  & $5.4261(5)$ & $5.3827(2)$ & $5.34024(6)$ &
                                  $5.29667(4)$ & $5.25038(5)$ \\
\hline   \end{tabular}  \vskip 5mm
Table~3: $Q=2$-quasiparticle energies at $\la=0.50$ for two values of
the momentum $P$ and varying chiral angles $\phi = \vphi$.
The values in the Table are estimates for the limit $\NIF$ obtained
from the values at $N=12$, $9$, $6$ and $N=12$, $10$, $8$ sites, respectively.
 \vskip 5mm   \end{table}
\par In order to illustrate that the integrable case is analogous
to the self-dual one for the parameters under consideration we also mention
two further values of ${\sf E}_2(P)$: For the {\em integrable} choice
of parameters $\la = 0.50$, $\vphi = 114^{\circ}$ one finds
${\sf E}_2(P= 2 \pi/3) = 5.683(2)$ and ${\sf E}_2(P= \pi) = 5.30608(3)$.
Note the similarity between these two values including their errors with
the corresponding ones in Table~3.          \par We conclude this Section
with the remark that for small $\la$ the behaviours of the self-dual
($\phi=\vphi$) and the integrable versions (\ref{in}) {\em have} to be
different, because in the integrable case for small $\la$ the non-hermitian
region comes close to the line $\sui$, whereas in the self-dual case little
special is happening for small $\la$ when moving in $\phi$ around $\phi=\ph$.

\section{Correlation functions}
In this Section we define and study a non-translationally invariant
correlation function for the operator $\Ga$ of the $\Zd$-chain numerically.
We shall use the definition
\BEQ \cg(x):= \frac{\LV\Ga_{x+1}^{+}\Ga_1\RV}{\LV v\rangle}
  \hs\hs 0\le x<\frac{N}{2}    \label{C}      \EEQ
where $\RV$ is the lowest eigenvector of the Hamiltonian in the sector
$Q=P=0$ which in the massive high-temperature phase is the ground state.
The definition (\ref{C}) ensures $\cg(-x)=\cg(x)^{*}$.
Non-translationally invariant correlation functions are difficult
to measure experimentally but for our purpose it will be crucial
to work really with $\Ga_{x+1}^{+} \Ga_1$ and {\it not}
to replace it by $N^{-1}\sum_{r=1}^{N} \Ga_{x+r}^{+} \Ga_r$,
because this would destroy the oscillation.
\par Because we are studying a massive phase, we expect an exponential decay
in addition to the oscillatory contribution. So a simple ansatz for
$\cg(x)$ may be made in terms of a complex correlation length or equivalently,
in terms of the real correlation length $\xi_\Ga$ and the oscillation
length $L$ defined through  \BEQ
\cg(x)=a e^{2\pi ix/L}e^{-x/\xi_{\Ga}}+(1-a)\delta_{x,0}. \label{ca}\EEQ

\par For the massive {\it low-temperature} phase a non-vanishing wave
vector has been predicted in \cite{geh} where also its critical
exponent was calculated from level crossings.
Perturbative calculations of the correlation function $\cg(x)$
for the massive low-temperature phase have been performed in \cite{hah}
and indications for an oscillatory contribution have been obtained.
Detailed perturbative studies of $\cg(x)$ in the massive
{\it high-temperature} phase will be presented in~\cite{per}.

\par We have noticed in \cite{gho} that in the high-temperature limit, the
dispersion curve of the $Q=1$-particle has its minimum at
$\Pmi=\RE\phi/3$, as it is obvious from the term of first order in $\la$
of $\Ee(P)$, see eq.\ (\ref{ee}). The value of $\Pmi$ decreases
with increasing $\la$: e.g.\ for $\phi=\vphi=\pi/2$ one finds from (\ref{Sz}),
(\ref{vrs}) that at $\la=0.901292$ $\Pmi\approx 0.225$ \footnote{More precise
parameter values for the transition to the IC phase for $\sui$ are:
\mbox{$\la=0.90129284\ldots$}\,,$\;\;\Pmi=0.22503375(3)$.} so that at this
$\la$,
$\Pmi$ has diminished by a factor of $\approx 2.33$ from its value at small
$\la$. In Sec.~2 (see Fig.~1) we have also noticed that for large $\la$
the third order formula (\ref{ee}) does not describe the dispersion curve
around the minimum.           \par The non-zero shift of the
minimum of the dispersion curve has the result that at $\phi\neq 0$,
already for $\la\ra 0$, the first excited state is no longer at
$P=0$. For a finite $N$-site chain with chain momentum $k=N P/2\pi$, the
lowest gap has $\Delta k=[N \Pmi/2\pi]$. This gives rise to a proliferation
of cross-overs in the first excited states with increasing $N$ which
in turn, via finite-size scaling arguments \cite{hgr}, indicates an
oscillating $\Delta P\neq 0$ correlation function. So we may expect
$\Pmi$ to be related to the oscillation length $L$ of the correlation
function.
\par For $\la\ra 0$ this relation is readily found using a high-temperature
expansion for the correlation function (\ref{C}). Calculating (\ref{C}) up
to second order in $\la$, we obtain
\BEQ \cg(x)=\de_{x,0} +\la\de_{x,1}\;\frac{e^{i\phi/3}}
{3\cos{(\pdr)}}+\frac{\la^2}{12\cos^2{(\pdr)}}\left\{\de_{x,1}\;e^{-\zpz}
+2\de_{x,2}\;e^{\zpz}\right\} +{\cal O}(\la^3).    \label{coo} \EEQ
In order to extract the complex correlation length from this formula to
lowest nontrivial order, we consider \BEQ    \frac{\cg(2)}{\cg(1)}=
\frac{e^{i\phi/3}}{2\cos{(\pdr)}}.  \label{cra}   \EEQ
On the other hand, from (\ref{ca}) we have   \BEQ \frac{\cg(2)}{\cg(1)}=
  \exp{\left(\frac{2\pi i}{L}-\frac{1}{\xi_\Ga}\right)}
,\label{crb} \EEQ so that, comparing (\ref{cra}) with (\ref{crb}) we
obtain \BEQ L\;\;\RE\phi=6\pi, \hs \xi_\Ga^{-1}=\frac{\IM\phi}{3}
 -\ln\frac{\la}{2\cos(\frac{\vphi}{3})}.  \label{LRe}  \EEQ
(We write $\RE\phi$, since e.g.\ in the integrable case (\ref{in})
 $\phi$ can be complex for small $\la$). Using the small-$\la$ expression
for $\Pmi$, the first of these equations can be rewritten as
a relation between the oscillation length
$L$ and the minimum $\Pmi$ of the dispersion relation (\ref{ee}):
\BEQ   \Pmi L = 2\pi,      \label{pl}    \EEQ
which is valid
at general $\phi$, $\vphi$ and small inverse temperatures $\la$.
\par In the following, we will show that with a finite-size numerical
calculation, using up to 13 sites, one can get useful information on the
oscillation length $L$ at various values of the parameters in the
high-temperature phase.
Since, as we have seen, $L$ is conceptually linked to $\Pmi$, we shall
conveniently give our results in terms of the product $\Pmi L$. It will
turn out that (\ref{pl}) seems to be satisfied approximately for a wide
range of parameters.
\par  In order to evaluate the correlation function (\ref{C}) numerically,
one needs the ground state of the Hamiltonian (\ref{H3}). The ground
state of the $\;\ZZ_3$-chain is easily obtained for general values
of the parameters (even in the non-hermitian case $\phi \in \CC$ \cite{yil})
using vector iteration up to $N=13$ sites. The more difficult
point is to calculate the matrix elements of $\Ga_{x+1}^{+} \Ga_1$,
because this operator does not conserve momentum and thus does
not leave a space of momentum eigenstates invariant. Table 4 shows
the correlation function $\cg(x)$ obtained in this manner for
$N=12$ and $N=13$ sites at two different points in the phase diagram:
first for $\la=0.5$ in the superintegrable case, and second, for $\la=0.25$
and the chiral angle $\vphi$ {\em above} the superintegrable line and
$\phi$ chosen complex (so that the Hamiltonian becomes non-hermitian)
satisfying the integrability condition (\ref{in}).
\begin{table}
\vskip 5mm   \noindent
\renewcommand{\arraystretch}{1.3}
\begin{tabular}{|c|cc|cc|}    \hline
\multicolumn{1}{|c}{ }&\multicolumn{2}{|c}{$\phi=\vphi=\kph,\hj \la=0.50$}&
\multicolumn{2}{|c|}{$\vphi=\kdv,\hj\phi=\pi+1.70005\,i,\hj\la=0.25$}\\ \hline
 $x$&{\it 12 sites}&{\it 13 sites}&{\it 12 sites}&{\it 13 sites}  \\  \hline
0&  1                &  1                 &  1   &  1  \\
1&$.18881+.07385i$ &$.18882+.07385i$ & $.06242+.10811i$ &$ .06242+.10811i$ \\
2&$.04587+.03967i$ &$.04588+.03967i$ & $-.01305+.02260i$&$-.01305+.02260i$ \\
3&$.01004+.01737i$ &$.01007+.01738i$ & $-.00668$        &$-.00667+.00000i$ \\
4&$.00126+.00679i$ &$.00132+.00684i$ & $-.00096-0.00167i$&$-.00096-.00165i$\\
5&$-.00056+.00224i$&$-.00043+.00242i$& $.00032-.00056 i$&$.00028-.00052i$  \\
6&$-.00080$        &$-.00063+.00058i$& $.00037$         &$.00022-.00005i$  \\
  \hline   \end{tabular}                \vskip 5mm
\par \noindent Table 4: Numerical results for the correlation
                     function $C_{\Ga}(x)$ at $N=12,13$ sites    \vskip 5mm
\end{table}
\par The excellent agreement between the calculations at $N=12$ and $N=13$
 sites shows that the contribution of boundary terms to the correlation
 functions is small. So we can use the $N=12-$values as a good
 approximation for the infinite chain limit as long as $x \le 6$.
\par We thus obtain six complex numerical values for $C_\Ga$ at each fixed
 point $(\la,\phi,\vphi)$ in the phase diagram considered. By a fitting
 procedure, we want to translate these six values into the three parameters
 $\xi_\Ga$, $L$ and $a$ of (\ref{ca}). We proceed as follows: First,
 we obtain $\xi_{\Ga}$ by calculating
 $\RE(\ln({C_\Ga(x)/C_\Ga(x+1) }))^{-1}$ and averaging
 over $x$. Next, the first zero of $\RE(e^{x/\xi_\Ga} C_\Ga(x))$
 is estimated by linear interpolation for two neighbouring
 values and $L/4$ is obtained by averaging. Finally,
 $a$ is chosen such that the difference
 \BEQ \RE C_\Ga(x)-a e^{-x/\xi_\Ga}\cos{(2\pi x/L)}  \label{rec}\EEQ
 is minimal for $x=1,2$. Table 5 collects our results for 12 choices of
 $\la$ and the chiral angles.
\par For the entries marked with a~`${}^{*}$' $\;\RE C_\Ga$ does not yet
 reach zero in the interval [0,6]. Here we calculated $L$ from the formula
\BEQ L=\kv \sum_{x=1}^4  \frac{2\pi}
           {\IM\left(\ln{(C_\Ga(x)/C_\Ga(x+1))}\right)}. \EEQ
The values of $\Pmi$ in Table 5 have been obtained by
first calculating $\Ee(P)$ numerically at $N=12$ sites and afterwards
minimizing the finite Fourier decomposition of $\left(\Ee(P)\right)^2$
 numerically - compare also Table 8 of \cite{gho}. These values are
 approximations with errors which are difficult to estimate.
\begin{table}
\vskip 5mm   \renewcommand{\arraystretch}{1.3}
\begin{tabular}{|c|c|c|c|c|c|c|c|}   \hline
$\vphi$ &$\phi$ &$\la$ &$\xi_\Ga$ &$L$ &$a$ & $\Pmi$& $L\Pmi/2\pi$
\\   \hline
$\kda$ & $\kda$ & $\kv$ & 0.53(3)& 18(2)& 0.65(3)  &0.341 & 1.0(1))     \\
$\kda$ & $\kda$ & $\kh$ & 0.8(2) &$25(1)^\ast$&    &0.283 & 1.14(6)     \\
$\kda$ & $\kda$ & $\kd$ & 1.8(6) &$50\pm23^\ast$&  &0.211 & 1.7(8)      \\
$\kph$ & $\kph$ & $\kv$ & 0.6(1) &14.05(6)& 0.44(8) &0.470 & 1.051(4)    \\
$\kph$ & $\kph$ & $\kh$ & 0.60(5)& 16(1)  & 0.59(3)  &0.401 & 1.07(8)     \\
$\kph$ & $\kph$ & $\kd$ & 1.3(2) &$29(6)^\ast$&      &0.189 & 0.9(2)\\
$\kdv$ & $\kdv$ & $\kv$ &0.62(8)& 10(1)& 0.4(2)& 0.746    &$1.18\pm 0.12$ \\
$\kdv$ & $\kdv$ & $\kh$ &1.2(6)&12(1)&0.4(1)&$0.685^\dagger$&$1.34\pm 0.14$\\
$\kdv$ & $\kdv$ & $\kd$ &1.4(1) & 16(1)& 0.65(3)&$0.602^\dagger$ &1.54(8)  \\
   \hline
$\kda$ &$-0.98942i$ &$\kv$& 0.7(2) & $\infty$ &    &0     &      \\
$\kda$ & 0.69919& $\kh$ & 1.1(3) &$49(7)^\ast$&    &0.159 & 1.2(2)\\
$\kdv$&$\pi+1.70004i$&$\kv$& 1.0(5) & 5.9(1) & 0.3(1) &$\kpd$& 0.99(2)  \\
\hline \end{tabular}     \vskip 5mm
\par \noindent Table 5: Parameters for the correlation function (\ref{ca})
calculated numerically for $N=12$ sites. In the first nine lines
we give data for the choice $\phi=\vphi$ of the chiral angles, where three
lines correspond to the superintegrable case $\sui$.
The bottom three lines use the integrable choice (\ref{in}).  \end{table}

\par The eighth and ninth line in Table~5 give situations close to,
and {\em in} the IC phase. In both cases $\Ee(P)$ partly
enters two-particle scattering states. For $\phi=\vphi=\kdv,~\la=\kd$
we are in the IC-phase with $\Ee(P)$ becoming negative. We nevertheless
define the correlation function for these cases with $\RV$ still denoting the
$Q=P=0-$ground state, although it is no longer the lowest state of (\ref{H3}).
Since in these cases the method explained above for estimating $\Pmi$
becomes impractical, here we proceeded differently: We first determine
the smallest energy gap in the $Q=1$ sector at $N=8,\ldots,12$ which appears
at the smallest value possible for the discrete
momentum, $P=2\pi/N$. Next, a polynomial interpolation between
these five values for the energy gap was minimized numerically.
In Table 5 we marked these results by `${}^{\dag}$'.
\par At the end of Table 5 we give three examples with integrable choices
of the chiral angles (\ref{in}). For $\vphi =\kda,~\la =\kh$,
the integrable Hamiltonian is still hermitian and behaves
very similarly to the the self-dual case. However, for small values of
$\la$ and $\phi\neq \pi/2$, one has to make one of the angles $\vphi$,
$\phi$ complex in order to satisfy (\ref{in}). We choose
$\phi$ complex. Below the superintegrable line $\sui$ one can
take $\phi$ purely imaginary, whereas above the superintegrable
line one has to admit a non-vanishing real part, e.g.\
$\phi-\pi \in i\,\RR$. With these choices
one can verify, using e.g.\ (\ref{ee}), that $\Ee(-P) = \Ee(P)^{*}$
for $\vphi < \ph$ and imaginary $\phi$ (for details, see \cite{yil}).
Therefore, it seems sensible to set $\Pmi=0$, and, if (\ref{pl})
holds at least approximately, we expect no
oscillation. Indeed, the complete correlation function $C_{\Ga}(x)$
is real for the integrable choice of parameters
$\vphi=\kda$, $\la=\kv$ which in view of (\ref{ca}), implies $L= \infty$.
For the integrable point $\vphi=\kdv,~\la=\kv$ above the superintegrable
line we have verified that (\ref{ee}) and a numerical evaluation of
$\Ee(P)$ at $N=12$ sites satisfy $\Ee(-P-\kpd) = \Ee(P-\kpd)^{*}$. Thus,
we set $\Pmi=\kpd$ which gives excellent agreement with the prediction
$L \Pmi = 2\pi$.
\par That the above procedures yield reasonable
fits is demonstrated by Fig.~6 which shows the stretched correlation
function $e^{x/\xi_{\Ga}} C_{\Ga}(x)$ obtained numerically for
$N=13$ sites at the integrable point $\vphi = \kdv$, $\la = \kv$ in
comparison to the fits. The errors have been estimated from the difference
between the results for $N=12$ and $N=13$ sites in Table 4. The agreement
for all $x$, not only in the real part but also in the imaginary part
is good. In particular, the oscillation is clearly visible.
\par The conjecture $L\Pmi=2\pi$ is well verified for all
values in Table 5, bearing in mind that we have ignored systematic errors.
This may apply also to the point $\phi=\vphi=\kdv, \la=\kd$ in the massless
IC phase, if we had underestimated the error in the determination of $L$.

\section{Conclusions}
\par In this paper we have presented further numerical evidence that the
 spectrum of the general $\;\ZZ_3$-spin quantum chain (\ref{H3}) in the
 massive high-temperature phase can be interpreted in terms of two
 fundamental quasiparticles and their scattering states.
Using duality \cite{hah} our results about the quasiparticle spectra
can be pulled over to the massive low-temperature phase.
\par The quasiparticle carrying $\;\ZZ_3$-charge $Q=2$ shows a quite peculiar
energy-mo\-men\-tum relation. It becomes unstable for a certain range of the
momentum, while exhibiting a remarkable stability against decay into
multi-particle states allowed by energy-momentum conservation in other ranges
of the momentum. Since we find this feature not only in the superintegrable
case, this generalizes the analytic result found by the
Stony-Brook group \cite{dkc} to cases for which no exact methods are known.
\par We also calculated the correlation function for the operator $\Ga$
using a numerical evaluation of the ground state of the $\;\ZZ_3$-model for
up to 13 sites. Although this approach is limited to short ranges, we are
able to estimate correlation lengths in the massive high-temperature
phase and can show that the correlation functions oscillate. We find
that the product of the oscillation length $L$ and the minimum of the
dispersion curve $\Pmi$ (which is non-zero due to parity violation)
satisfies $L\Pmi\approx 2\pi$.
\vskip 1 cm \par \noindent {\large {\bf Acknowledgement} }
We like to thank Fabian E{\ss}ler and Karim Yildirim for useful discussions.

\newpage
\par \noindent {\Large {\bf Figure captions} }
\begin{description}
\item{Fig.\ 1:} Momentum dependence of the spectrum of the superintegrable
  $\Zd$-Hamiltonian in the $Q=1$-sector for $\la=0.90$. Symbols: Finite
  chain data, dashed curve: exact lowest level from \cite{acp}, point-dashed
  curve: $3rd$-order perturbation in $\la$ according to eq.~(\ref{ee}).
\item{Fig.\ 2:} Spectrum of the $\Zd$-Hamiltonian at $\phi = \vphi =
  \pv$, $\la = 0.10$. All three charge sectors have been plotted in the same
  figure. The dots indicate eigenvalues for $N=12$ sites whereas the lines
  are obtained from the $3rd$-order perturbative expansions (\ref{ee}) and
  (\ref{ez}). We have limited ourselves to show for each $(Q, P)$ the nine
  lowest eigenvalues only. One observes that the eigenvalues nicely group
  within the band boundaries expected according to (\ref{ban}). For $Q=1$ and
  $Q=2$ the first levels building up the three-particle bands are seen for
  $E>8.5$. For a discussion of the four $Q=0$-points at $P=\pi$ outside the
  $(\Ee+\Ez)$-band, see the text.
\item{Fig.\ 3:} Same as Fig.\ 2 but for $\phi=\vphi= \kda$, $\la=0.25 $. Note
  that all three charge sectors have been plotted in the same figure. Thus,
  seemingly overlapping bands belong to different charge sectors and so
  are no real level crossings. The dots indicate eigenvalues for
  $N=4,\ldots,12$ sites. The curves show the perturbative results for
  $\Ee(P)$ (full curve), $\Es_{Q=2}^\pm(P)$ (see eq.~(\ref{ban}), long dashes)
  and $\Ez(P)$ (fine dashes), according to (\ref{ee}),~(\ref{ez}).
  \par For some range of $P$ the $Q=2$ excitation
  enters the energy band of two $Q=1$ particles. In this region, the
  $Q=2$ "particle" is difficult to trace, although outside this region
  the convergence in $N$ is very fast. This indicates that the $Q=2$
  fundamental excitation is not stable for all momenta $P$.
 \item{Fig.\ 4:} Spectrum of the superintegrable $\Zd$-Hamiltonian at
   $\sui$, $\la=0.5$. All three charge sectors have been plotted in the
   same figure but only $N=12$-finite-size data are shown.
   The dashed curves show the analytic result of \cite{dkc}, eq.~(\ref{Sz})
   for $2\Ee(P/2)$ (long dashes) and $\Es_{Q=2}^\pm(P)$ (fine dashes).
   The full curve is the analytic result for $\Ez(P)$. Note that the $Q=2$
   fundamental particle exists only for $P \ge \frac{2\pi}{3}$.
\item{Fig.\ 5:} Part of the spectrum in the $Q=2$-sector at $\la=0.50$ and
   fixed momentum $P=2\pi/3$ for $N=6, 9, 12$ sites, as a function of the
   chiral angle $\phi=\vphi$ (self-dual choice). On sees that just one level
   from $N=6, 9, 12$ each moves very closely together with their
   companions from the other $N$ when $\phi$ varies. We take this as
   evidence for an exponential convergence of this level in $N$,
   characteristic for a single-particle state. In the upper right corner
   there are more $N=12$-levels which have not been drawn since we limited
   ourselves to calculate for $N=12$ no more than 29 levels.
\item{Fig.\ 6:} Numerical values for the correlation function $C_{\Ga}(x)$
   at $N=13$ sites stretched by $e^{x/\xi_\Ga}$
   in comparison to the fit (\ref{ca}) at $\phi = \kdv$,
   $\la =\kv$, $\phi=\pi +1.70004i$. The error bars are given by
   the differences between $N=12$ and $N=13$ sites.
   The parameters used for the fit are $\xi_\Ga = 0.77$, $L = 6$
   and $a=0.35$. One observes that the agreement with these fits is good and
   the oscillatory contribution to $C_{\Ga}(x)$ is clearly visible.
\end{description}
\end{document}